\documentclass{iau} 
\usepackage{graphicx}
\usepackage{amsmath,amsxtra}

\usepackage{multicol}
\usepackage{etoolbox}
\usepackage{relsize}
\patchcmd{\thebibliography}
{\list}
{\begin{multicols}{2}\smaller\list}
	{}
	{}
	\appto{\endthebibliography}{\end{multicols}}

\title[Molecular gas in the immediate vicinity of Sgr A*] 
{Molecular gas in the immediate vicinity of Sgr A* seen with ALMA}

\author[L. Moser, \'A. S\'anchez-Monge, A. Eckart, et al.]   
{Lydia Moser$^{1,2,3}$,
 \'Alvaro S\'anchez-Monge$^2$, 
 Andreas Eckart$^{2,3}$,
 Miguel A. Requena-Torres$^4$,
 Macarena Garc\'ia-Marin$^5$,
 Devaky Kunneriath$^6$,
 Anton Zensus$^{2,3}$,
 Silke Britzen$^3$,
 Nadeen Sabha$^2$,
 Banafsheh Shahzamanian$^{2,3}$,	
 Abhijeet Borkar$^{2,7}$, \and 
Sebastian Fischer$^8$}

\affiliation{$^1$Argelander-Institut f\"ur Astronomie, University of Bonn, Auf dem H\"ugel 71, 53121 Bonn, Germany --- email: {\tt lmoser@uni-bonn.de} \\[\affilskip]
$^2$I. Physikalisches Institut, Universit\"at zu K\"oln, Z\"ulpicher Str. 77, 50937 K\"oln, Germany \\[\affilskip]
$^3$Max-Planck-Institut f\"ur Radioastronomie, Auf dem H\"ugel 69, 53121 Bonn, Germany \\[\affilskip]
$^4$Space Telescope Science Institute, 3700 San Martin Dr., Baltimore, 21218 MD, USA \\[\affilskip]
$^5$European Space Agency, 3700 San Martin Drive, Baltimore, 21218 MD, USA \\[\affilskip]
$^6$National Radio Astronomy Observatory, 520 Edgemont Road, Charlottesville 22903, USA \\[\affilskip]
$^7$Astronomical Institute, Academy of Sciences, Bo{\v c}n\'i II 1401, CZ-14131 Prague,\,Czech\,Republic \\[\affilskip]
$^8$German Aerospace Center (DLR), K\"onigswinterer Str. 522-524, 53227 Bonn, Germany \\[\affilskip]
}

\pubyear{2016}
\volume{322}  
\jname{Multi-Messenger Astrophysics of the Galactic Centre}
\editors{S. Longmore, G. Bicknell \& R. Crocker, eds.}
\begin{document}

\maketitle

\begin{abstract}
We report serendipitous detections of line emission with ALMA in band 3, 6, and 7 in the central parsec of the Galactic center at an up to now highest resolution ($<$0.7$''$). Among the highlights are the very first and highly resolved images of sub-mm molecular emission of CS, H$^{13}$CO$^+$, HC$_3$N, SiO, SO, C$_2$H, and CH$_3$OH in the immediate vicinity ($\sim$1$''$ in projection) of Sgr A* and in the circumnuclear disk (CND).
The central association (CA) of molecular clouds shows three times higher CS/X (X: any other observed molecule) luminosity ratios than the CND suggesting a combination of higher excitation - by a temperature gradient and/or IR-pumping - and abundance enhancement due to UV- and/or X-ray emission. We conclude that the CA is closer to the center than the CND is and could be an infalling clump consisting of denser cloud cores embedded in diffuse gas.
Moreover, we identified further regions in and outside the CND that are ideally suited for future studies in the scope of hot/cold core and extreme PDR/XDR chemistry and consequent star formation in the central few parsecs. 


\end{abstract}

\firstsection 

\begin{figure}[htb]
	\begin{center}
		\includegraphics[trim = 10mm 40mm 0mm 40mm, clip, width=\textwidth]{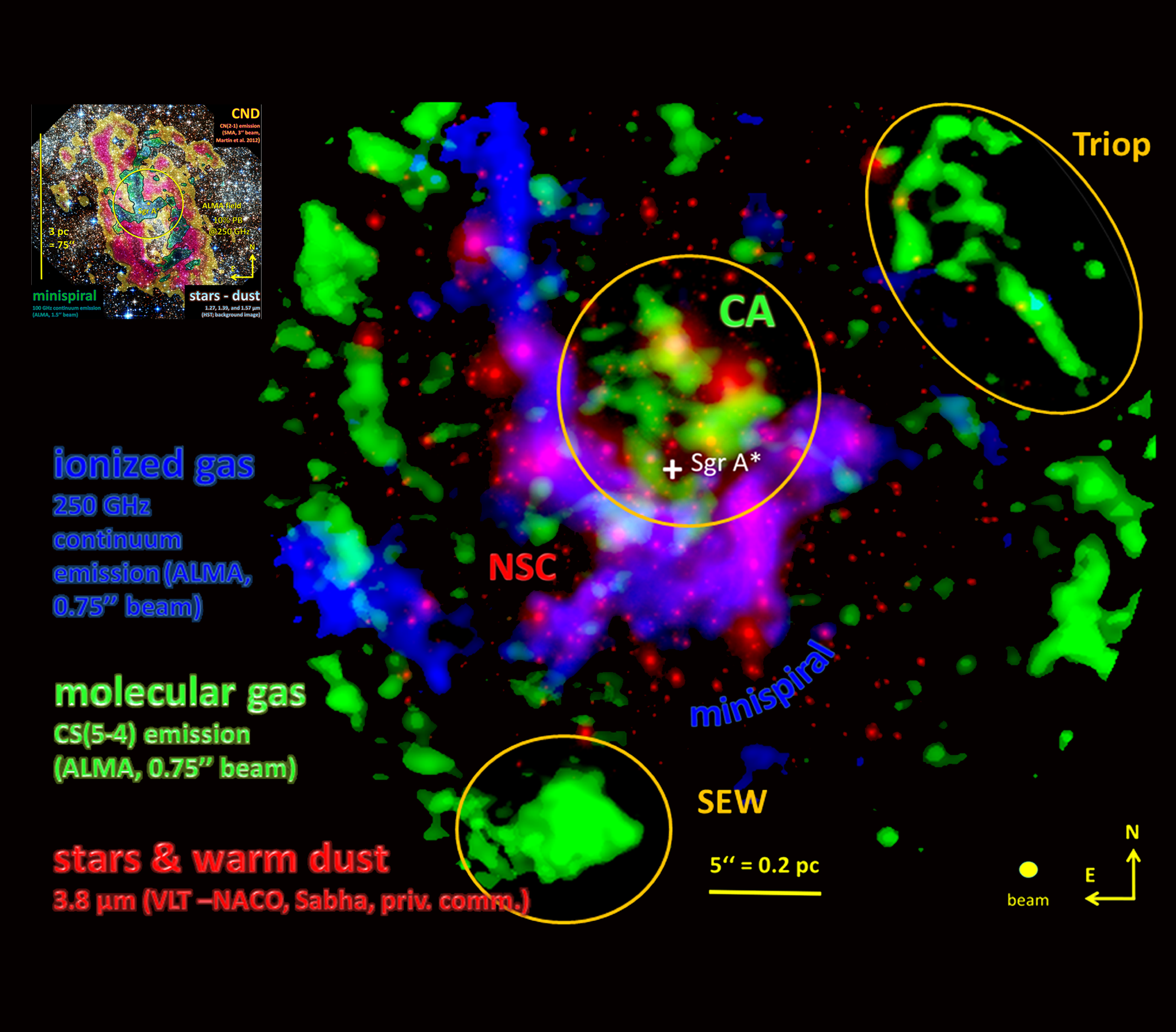} 
		\caption{
			\textit{Upper left corner:} The CND in CN(2--1) emission (\cite{Martin2012}) in yellow-red (extended grey structure), the minispiral in 100 GHz continuum emission (this data) in green-blue (extended grey structure within dark contour) on an Hubble Space Telescope (HST) near-infrared (NIR) composite image (HST archive). The yellow (light grey) ring denotes the region (10\% ALMA primary beam sensitivity at 250 GHz) visible in the main figure. 
			\textit{Main figure:} The inner 40$''$ seen with ALMA - Green (light grey) shows the molecular gas traced by CS(5--4) in the center and at the inner edges of the CND, blue (dark grey) shows the continuum emission of the ionized 'minispiral' gas at 250 GHz. Stars and dust detected in the NIR L' band (Sabha, private communication) are shown in red (mid grey point sources; for dust see online version). For a detailed view on scales and fluxes see \cite{Moser2016}.
		}
		\label{fig1}
	\end{center}
\end{figure}

\section{Introduction}

Sagittarius A* (Sgr A*) at the center (distance: $\sim$8 kpc; e.g. \cite{schoedel2002}) of our Galaxy is the closest supermassive black hole (SMBH) for which we can study its environment on scales less than 1000 AU ($\sim$4 mpc). 
The nuclear stellar cluster (NSC; Fig. \ref{fig1}) engulfing Sgr A* comprises several extremely young stars (e.g. \cite{Krabbe1995}) which raises the question on the star forming potential of the molecular gas in the cavity and at a ring at 1.5 pc distance - the circumnuclear disk (CND; Fig. \ref{fig1}) - and its transport to the center.

In \cite{Moser2016}, we present Atacama Large Millimeter/ submillimeter Array (ALMA) data (project: 2011.0.00887.S) of the central parsecs of our Galaxy at 100, 250, and 340 GHz and at a to date highest spatial resolution of 1.5$''$, 0.75$''$, and 0.5$''$ ($\sim$0.02 pc), respectively. 
Here, we summarize our most important findings regarding the molecular gas emission.

\section{Central association  of gas clumps}

Previous interferometric observations (\cite{Montero2009,Martin2012}) have indicated the presence of molecular gas close (few arcseconds in projection) to Sgr A*. 
Our ALMA data confirms this and gives the first high spatial resolution ($\lesssim$ 0.75$''$) probe of this region in CS(5--4) (brightest; Fig. \ref{fig1}), H$^{13}$CO$^+$(3--2), HC$_3$N(27--26), SiO(6--5), SO(6--5), C$_2$H(3--2), and CH$_3$OH(7--6). The clumpy structure of this central association (CA) reaches less than 1$''$ to the SMBH and covers velocities from about -80 km s$^{-1}$ southeast to +80 km s$^{-1}$ northwest of Sgr A*. 

Since the line excitation is governed by several, for this region unconstrained factors, we cannot derive any density or temperature ranges of the molecular gas. Instead we can track changes in the interstellar matter (ISM) properties by the help of line luminosity ratios. We find that the CA shows three times higher CS/X (X: any other observed molecule) luminosity ratios than the CND. 
The NSC offers a variety of explanations: First, it heats the environment, hence a temperature gradient is likely. Second, it has a high infrared (IR) output, so that ro-vibrational transitions can be strongly excited and result in an overpopulation of the rotational levels (IR-pumping of CS, HCN, SiO; \cite{Carroll1981}). 
Third, the NSC produces a strong UV- and X-ray field likely to change the chemical composition as seen in (edges of) Photo-Dissociation Regions (PDR)/X-ray Dominated Regions (XDR). Forth, shock chemistry may occur when the CA gas collides with the stellar winds.
No single cause can explain the emission behaviour of each molecule satisfyingly, so that most likely a combination of them is at work.
\cite{Goto2014} obtained a temperature and density estimated from H$_3^+$ and CO spectra towards two positions within the CA of  $T_\textrm{k} \sim$ 300 K and $n_\textrm{H$_2$}$ $\geq$ $10^4$ cm$^{-3}$, i.e. the molecular gas is warmer and less dense than the CND (\cite{RT2012,Mills2013}).
This all suggests that the CA is closer to the center than the CND is.

In terms of its velocity, it might be related to the  \textit{triop} (Fig. \ref{fig1}) - a filamentary feature in the CND to the Northwest of Sgr A* between 20 - 120 km s$^{-1}$. Another explanation is offered by a streamer detected in OH absorption by \cite{Karlsson2015}: This feature extends inside and along the southwestern CND (projection!) to the north and heads to the center from the southwest. Its compact "head" has a mass of 65 $M_\odot$ and is consistent in shape, position and velocity distribution with the CA. 
We estimated the mass of the CA analogue to \cite[Karlsson et al. (2015]{Karlsson2015}, i.e. a homogeneous ellipsoid with $n_\textrm{H$_2$}$ $\sim$ $10^5$ cm$^{-3}$) and obtain 5 $M_\odot$. The density traced by the OH transition could easily be a factor of 10 less than assumed by them, since OH can already be excited at much lower densities. Hence, the masses are roughly consistent. All in all, the OH steamer corroborates 
our scenario of an infalling clump consisting of denser cloud cores that are embedded in diffuse gas.

\section{Triop and the SEW cloud}

Despite its thin structure, the \textit{triop} (Fig. \ref{fig1}) in CND northwest of Sgr A* shows comparably bright emission in all molecules in this as well as in previous observations (e.g. \cite{Christopher2005,Montero2009,Martin2012}). The luminosity ratios we derived correspond to the CND average. The range of detected transitions includes complex molecules, UV-sensitive species, and high energy transitions (i.e. HC$_3$N(27--26) with $E_u/k_B\sim$165 K). Furthermore, the \textit{triop} harbours the closest (to Sgr A*) class I methanol maser (44 GHz; \cite{YZ2008,Sjouwerman2010}). 
With all these properties, the \textit{triop} could show signs of an early phase of star formation/hot core.

In contrast to this, the bright compact cloud at the western tip of the Southern Extension of the CND (SEW; Fig. \ref{fig1}) shows line ratios similar to the CA, but not as high. Either we are witnessing a slightly attenuated impact of the NSC on the gas in the SEW or there is another source of the PDR/XDR or a shock.

\begin{figure}[htb]
	\begin{center}
		\includegraphics[trim = 0mm 55mm 2mm 41mm, clip, width=\textwidth]{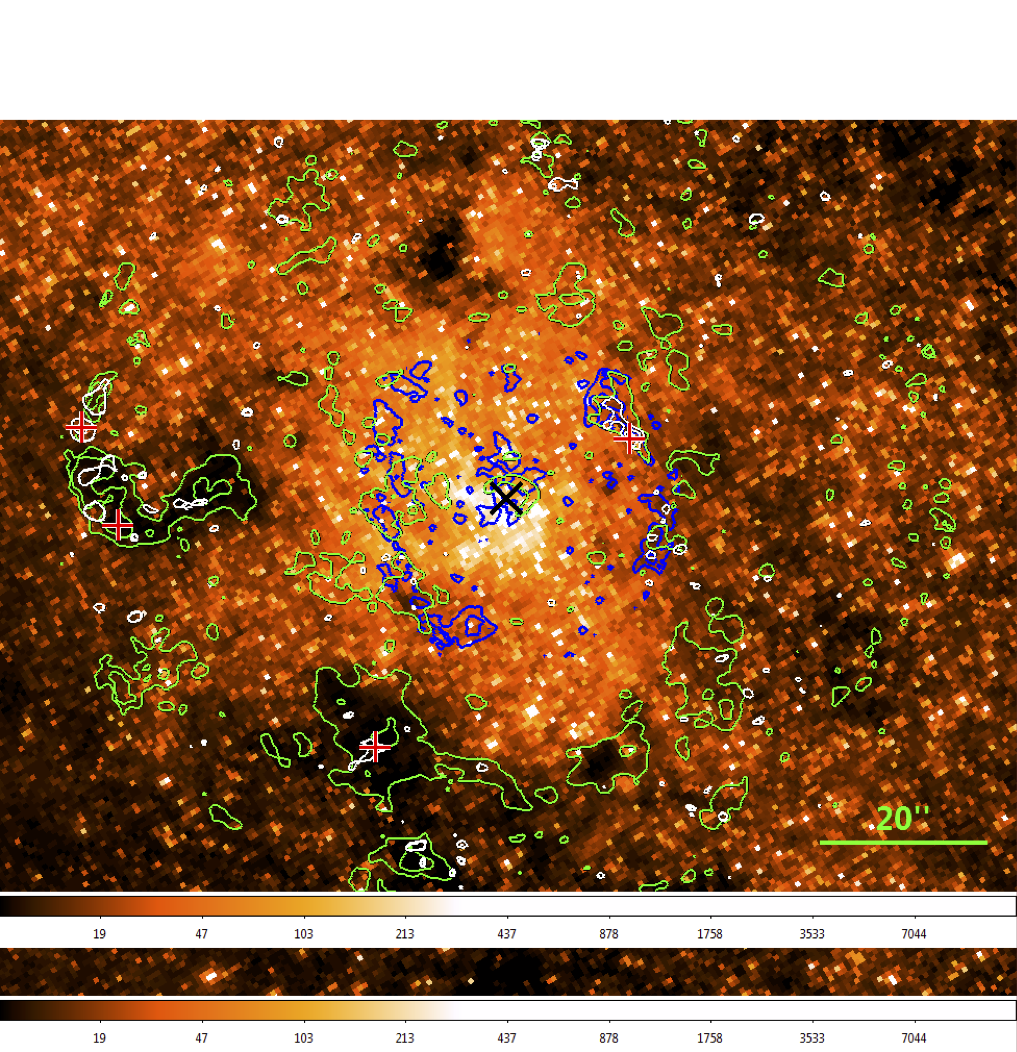} 
		\caption{
Schematic view on the central $\lesssim$ 5pc: N$_2$H$^+$(1--0) (green/light grey), CH$_3$OH(8--7) (white), and CS(5--4) (blue/dark grey) contours on a NIR HST (1.87$\mu$m) image (HST archive). 
Red/grey crosses in the \textit{triop} and the two large IRDCs to the (south-)east mark the class I methanol masers and the black cross Sgr A*. North is up, east to the left. For a detailed view on scales and fluxes see \cite{Moser2016}.
		}
		\label{fig2}
	\end{center}
\end{figure}

\section{Infrared-dark clouds}

In \cite{Moser2014}, we reported the detection of N$_2$H$^+$ and CH$_3$OH in the prominent infrared-dark clouds (IRDCs) east of the CND. The high resolution of this ALMA data set shows how closely the N$_2$H$^+$ outlines these IRDCs (Fig. \ref{fig2}). The presence of both, a tracer of cold ($T_\textrm{k} \sim 20$ K; \cite{Vasyunina2012}), quiescent, dense gas and the high opaqueness in the IR, indicates suitable conditions for prestellar (cold) cores and earliest stages of star formation. This could be supported by the class I methanol maser (36 \& 44 GHz; \cite{YZ2008,Sjouwerman2010,Pihlstrom2011}) in these regions (Fig. \ref{fig2}). However, the masers might rather be related to collisions between the giant molecular clouds (GMCs) in the east with the Sgr A East supernova remnant (SNR) shell as also the widespread SiO emission around these maser locations suggests (e.g. \cite[Moser et al. 2014]{Moser2014}).

\end{document}